# Hybrid physics-AI outperforms numerical weather prediction for extreme precipitation nowcasting.

Puja Das,[a]  August Posch,[b]  Nathan Barber,[c]  Michael Hicks,[c]  Thomas J. Vandal,[d]  Kate Duffy,[d] Debjani Singh,[e]  Katie van Werkhoven,[f]  Auroop R. Ganguly,[a,b,g]

[a] *Sustainability and Data Sciences Laboratory, Northeastern University, Boston, MA, USA*
[b] *The Institute for Experiential AI and Roux Institute, Northeastern University, Boston, MA, USA*
[c] *Tennessee Valley Authority, Knoxville, TN, USA*
[d] *Zeus AI, Cambridge, MA, USA*
[e] *Environmental Sciences Division, Oak Ridge National Laboratory, TN, USA*
[f] *Research Triangle Institute, Raleigh, NC, USA*
[g] *Pacific Northwest National Laboratory, Richland, WA, USA*

*Corresponding author*: Auroop R. Ganguly, a.ganguly@northeastern.edu

ABSTRACT: Precipitation nowcasting, critical for flood emergency and river management, has remained challenging for decades, although recent developments in deep generative modeling (DGM) suggest the possibility of improvements. River management centers, such as the Tennessee Valley Authority, have been using Numerical Weather Prediction (NWP) models for nowcasting but have struggled with missed detections even from best-in-class NWP models. While decades of prior research achieved limited improvements beyond advection and localized evolution, recent attempts have shown progress from physics-free machine learning (ML) methods and even greater improvements from physics-embedded ML approaches. Developers of DGM for nowcasting have compared their approaches with optical flow (a variant of advection) and meteorologists' judgment but not with NWP models. Further, they have not conducted independent co-evaluations with water resources and river managers. Here, we show that the state-of-the-art physics-embedded deep generative model, specifically NowcastNet, outperforms the High-Resolution Rapid Refresh (HRRR) model, the latest generation of NWP, along with advection and persistence, especially for heavy precipitation events. For grid-cell extremes over 16 mm/h, NowcastNet demonstrated a median critical success index (CSI) of 0.30, compared with a median CSI of 0.04 for HRRR. However, despite hydrologically relevant improvements in point-by-point forecasts from Nowcast-Net, caveats include the overestimation of spatially aggregated precipitation over longer lead times. Our co-evaluation with ML developers, hydrologists, and river managers suggests the possibility of improved flood emergency response and hydropower management.

# 1. Introduction

Flooding, a prevalent weather hazard, impacts numerous regions globally, causing economic damage and disruption each year. In the United States alone, between 1980 and 2019, flooding resulted in losses totaling $146.5 billion and claimed the lives of 555 individuals, as reported by NOAA's National Centers for Environmental Information[1]. The impacts of floods and especially flash floods extend beyond immediate human and infrastructural losses to encompass critical infrastructure such as hydropower operations and dam management[2]. In the Southeastern United States (SEUS), flash floods are a major concern due to their sudden and severe nature[3;4]. The Tennessee Valley Authority (TVA), which manages the Tennessee River system in Tennessee and six surrounding states in the SEUS, has to often deal with flash floods, primarily triggered by mesoscale storms with embedded convection (MEC). A noteworthy example is the devastating flood in middle Tennessee, in August 2021, which resulted in the loss of 20 lives and more than $100 million in property damage[5]. Similarly, the Damodar Valley Corporation, modeled after the TVA[6], manages the Damodar River in West Bengal, India, and has been struggling with unpredictable floods[7] despite infrastructural advancements. Other examples of river management centers dealing with deadly flash floods includes the Società Adriatica di Elettricità in Italy (1963 Vajont Dam failure, 2000 people killed)[8], the Nile River Basin Authority in Egypt (2015 Alexandria and Nile Delta floods, 17 deaths[9]), and the Kerala Water Resources Department in India (2018 flood in Kerala, 400 deaths[10]). To address the challenges posed by emergency flash flood management, short-term Quantitative Precipitation Forecasts (QPF) serve as vital tools by driving the hydrologic and hydraulic models which predict runoff and flooding downstream[11;12]. Traditional forecasting methods have employed persistence, advection of radar echoes[13], Numerical Weather Prediction (NWP) models[14], and data-driven extrapolation-based methods[15], either individually or in combination[16]. Although short-term QPFs offer well-documented advantages, this field has long been acknowledged as one of the most challenging in hydrometeorology. Even leading NWP models often struggle to accurately predict extreme precipitation events, prompting organizations like the TVA to replace top-tier NWP models with coarser alternatives. However, in recent years, with advancements of machine learning, studies have demonstrated that deep learning methods can surpass traditional approaches like persistence, advection, and optical flow[16–19].

Current machine learning methods treat forecasting as an image-to-image translation problem, employing computer vision tools to generate nowcasts[20]. The latest development in such physics-free nowcasting approaches comes from Google DeepMind[21]. Their physics-free AI model, known as Deep Generative Model of Rainfall (DGMR), is trained on historical weather data and can rapidly analyze patterns and make predictions without explicit knowledge of atmospheric physics. However, while DGMR offered accurate forecasts in comparison to previous methods, it struggled to accurately predict extreme precipitation events[22]. A more recent study improved nowcasting of extreme precipitation by combining physical-evolution schemes such as the conservation of mass for precipitation fields over time and space and conditional-learning methods into a neural-network framework called NowcastNet[22]. It addresses both advective and convective processes, which was previously deemed challenging in DGMR.

In this study, we assess the performance of the state-of-the-art physics-conditioned deep generative model in predicting precipitation patterns during record-breaking flood events as well as heavy precipitation events in the Tennessee Valley. Due to its extreme storms and extensively dammed rivers, the Tennessee Valley is a critical focus area for evaluating NowcastNet's effectiveness in flood prediction and disaster management. In this study, we evaluate the following methods:

- NowcastNet[22], state-of-the-art physics embedded DGM, provides forecasts at 10 min interval for 3 hours at 1 km resolution. NowcastNet merges convective-scale details observed through radar data with mesoscale patterns dictated by physical laws into a neural-network framework.

- High Resolution Rapid Refresh (HRRR)[23;24], state-of-the-art NWP model, developed by NOAA, provides hourly forecasts at 3 km resolution utilizing complex physics based equations and data assimilation.

- Baseline Approaches:

    - Advection or Optical Flow, represented by the PySTEPS[25] algorithm, which uses an advection scheme influenced by the continuity equation. It predicts future motion fields and intensity residuals by iteratively advecting past radar data.

    - Persistence, which assumes precipitation intensity and location will remain the same over increasing lead time.

While developers of physics free or physics-conditioned deep generative models of nowcasting, have compared their approaches with optical flow in terms of skill scores as well as judgment of meteorologists, they have not compared with NWP models and did not do independent evaluations for hydrologic use. This study compares NowcastNet with HRRR, which is widely used in many river management centers but has not previously been evaluated against NowcastNet. Accurate nowcasting impacts integral areas in hydrology such as river management, dam operations, and flash flood prediction, which affect the lives and property of human directly. So, we are evaluating the model with a wider set of metrics. Firstly, skill scores help in anticipating events, which is useful for emergency alerts such as issuing a flood warning. However, for precise management and response, such as predicting the exact flood levels or determining the exact volume of water to release from a dam, evaluation of continuous metrics is also needed. This approach ensures that the model's performance aligns with practical applications in flood forecasting, enhancing the reliability of predictions and reducing the risk of false alarms or misses that could lead to inappropriate responses in critical situations.

## 2. Results

*a. Tennessee Valley Authority Case Study*

The TVA plays a pivotal role in flood control, navigation, power generation, water supply, water quality maintenance, and recreation across the Tennessee River system in the Southeastern US and the Appalachian region. They manage a vast river network spanning approximately 640 miles and encompassing around 40,000 square miles of watershed. TVA manages 49 dams, 29 of which produce hydroelectricity, and they serve electricity to 153 local power companies serving more than 10 million people. Moreover, with strategically constructed dams and reservoirs along major river systems like the Tennessee River, TVA regulates water flow to mitigate flood risks during heavy rainfall and storm events. In Figure 1, the operating area of TVA is shown with the locations of key electricity generating facilities. In the Tennessee Valley, floods are primarily triggered by mesoscale storms with embedded convection (MEC), mid-latitude cyclones (MLC), and tropical storm remnants (TSR), either individually or in combination. Despite advancements in weather prediction models, several flood instances revealed limitations in forecasting intensity and location accurately. A pertinent case study is the devastating August 2021 flood in Waverly,

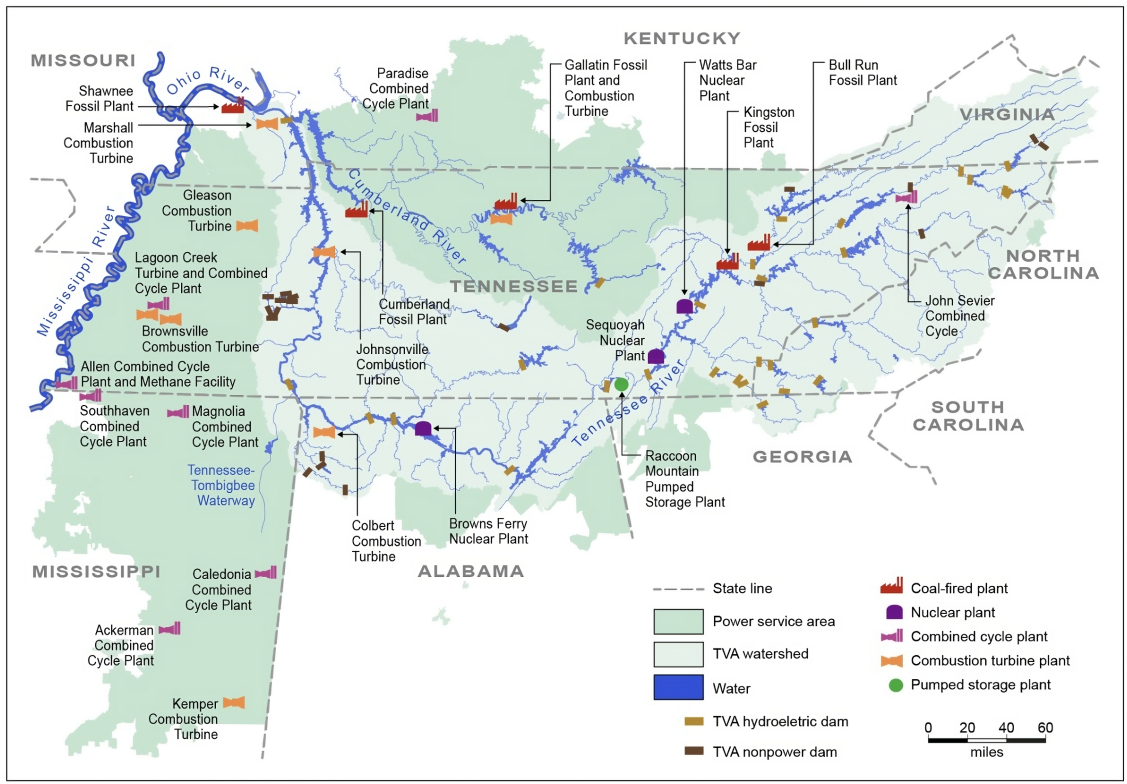

FIG. 1. Map depicting the Tennessee Valley Authority (TVA) service area and the locations of key electricity generating assets within the region, as reported by the Government Accountability Office (GAO) in their 2023 Report to Congressional Requesters[26]. The figure provides an overview of the geographical coverage of TVA's operations and highlights the distribution of major power generation facilities.

Tennessee. The flood event was triggered by unprecedented rainfall and the deluge, attributed to a complex interplay of meteorological phenomena, highlighted vulnerabilities in flood preparedness and response mechanisms. Despite prior warnings issued by the National Weather Service, the rapid onset of the flood prevented timely evacuation efforts, exacerbating the impact on residents. Meteorological observations indicated an abundance of atmospheric moisture, coupled with the convergence of a mid-level warm front and a stationary front over West Tennessee, creating conditions conducive to intense precipitation and subsequent flooding. The mesoscale convective system responsible for the event showcased the heightened vulnerability to extreme weather events, emphasizing the imperative for robust flood management strategies and precise forecasting methods to mitigate future risks. While this event occurred in an unregulated part of the basin, it underscores the potential for similar catastrophic events across the TVA region. TVA holds its dam reservoirs at a high water level in the summer, as part of its multi-objective optimization, including recreation and seasonal electricity demand. These elevated water levels would have constrained the time available for emergency response if the event had happened in a regulated section of the system. Thus, accurate forecasts would have been crucial in managing or mitigating the flood impact, emphasizing the significance of timely predictions. Therefore, the Waverly flood event emphasizes the intricate relationship between meteorological dynamics and human vulnerability, prompting TVA to prioritize high-quality hourly forecasts and consistent predictions for extreme events.

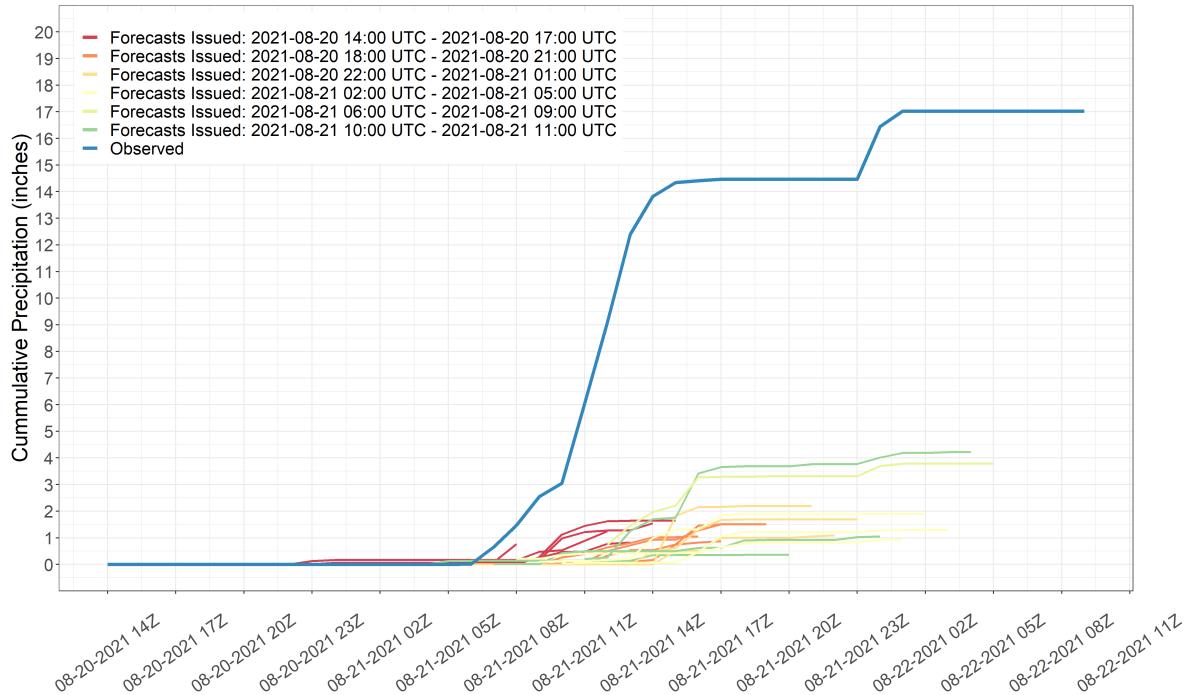

Fig. 2. Comparison of accumulated precipitation forecasts (in inches) from High-Resolution Rapid Refresh (HRRR) model samples at the McEwen Precipitation gauge with Observation (Blue Line) during the Waverly event on August 21, 2021, displayed at Coordinated Universal Time (UTC). The McEwen precipitation accumulation data utilized in this comparison is derived from TVA-collected data. The plot illustrates the discrepancy between the accumulated precipitation forecasts and the actual observations at the McEwen gauge, providing insights into the forecast bias.

Following the Waverly event, questions arose regarding the effectiveness of the HRRR model, used by TVA and other agencies to predict weather patterns and assess flood risks. While the HRRR model provided valuable insights into typical weather conditions, its performance during the Waverly event cast doubt on its reliability during extreme weather events. Here, Figure 2 shows the performance of the HRRR model during the Waverly event on August 21, 2021, displayed in Coordinated Universal Time (UTC). The figure reveals the disparity between the predicted accumulated precipitation and the observed values at the McEwen gauge, shedding light on the forecast bias. The McEwen precipitation accumulation data utilized in this comparison is derived from TVA-collected data. Around 11:00 UTC, when the actual rainfall accumulation reached a cumulative 13 inches, the HRRR forecasted only 2 inches. Similarly, despite a total rainfall of 17 inches throughout the day in McEwen, the HRRR model predicted only 4 inches. More information about other heavy precipitation events and HRRR's failure to accurately forecast the events are given in supplementary information section A. Beyond assessing the Waverly event, we also considered thirty additional extreme precipitation events (more than 30 mm/hr) occurring between January 2021 and April 2024, specifically within the TVA area. The list of the events is given in supplementary section (Table S1). These events are selected based on the catastrophic impacts they had in the TVA region.

## b. Performance of physics conditioned deep generative model: NowcastNet

The performance of the NowcastNet model during the Waverly event (August 21, 2021) is evaluated within the TVA area. Multi Radar Multi Sensor (MRMS) data are used as reference observations. MRMS provides estimated precipitation rate maps at 0.01° (1 km) spatial resolution and 2 min temporal resolution. Developed by NOAA's National Severe Storms Laboratory (NSSL), the dataset incorporates data from approximately 180 operational US WSR-88D weather radars and model analyses to produce gridded precipitation products[27]. The steps to apply the NowcastNet model in the TVA region are described in supplementary information section B. Figure 3 presents precipitation predictions starting from 9:00 UTC ($T + 1h$) until 11:00 UTC ($T + 3h$) from both the NowcastNet model and HRRR forecasts, along with the PSD performance metric.

The Waverly event is characterized by its extreme precipitation, which stemmed from a mesoscale convective system, a collection of thunderstorms. Capturing extreme precipitation at convective scales is challenging due to the rapid development, intensity, and localized nature of convective storms. Despite the challenges, NowcastNet predicted the hotspots of extreme precipitation locations of more than 30 mm/h more accurately than HRRR.

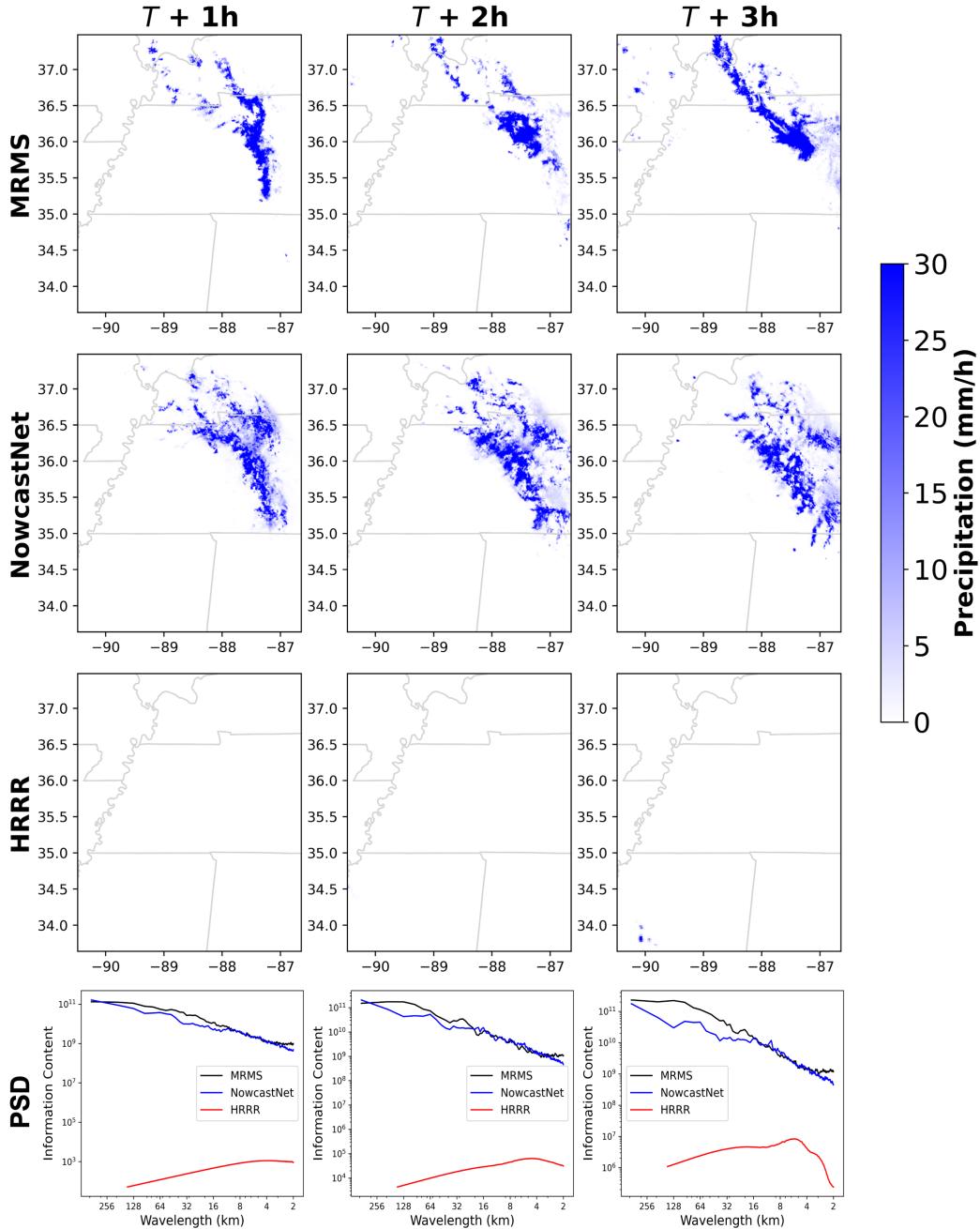

FIG. 3. Precipitation forecasts (in mm/h) from NowcastNet (1 km spatial resolution) and HRRR (3 km spatial resolution) at different lead times ($T$+1h, $T$+2h, and $T$+3h) with MRMS QPE[27] for the Waverly flood event on August 21, 2021 ($T$ = 8:00 UTC) within the TVA area. The precipitation images cover a spatial extent of 384 km × 384 km. The base map shows US state boundaries. NowcastNet predicts the MRMS precipitation patterns much more closely than HRRR does, in terms of the spatial distribution and intensity of the precipitation. The last row depicts the PSD at different wavelengths at different lead times ($T$+1h, $T$+2h, and $T$+3h).

For the 3-hour predictions, NowcastNet is capable of forecasting the trajectory of the thin line of convective precipitation, whereas HRRR could not predict the heavy precipitation at all. However, as the lead time increases, the NowcastNet model exhibits broader areas of light precipitation, leading to challenges in precisely identifying the exact location of precipitation. Additionally, there was a tendency for the model to overestimate precipitation over a broader area, so the discrepancy between observed and predicted precipitation also increases.

One success in the model's performance is that even at longer lead times, NowcastNet displays the right amount of detail at most spatial scales. The spatial power spectral density (PSD) reveals strength of a signal as a function of spatial scale, and for this case study the PSD curve of the forecast matches the PSD curve of the MRMS for wavelengths 4 km to 16 km, and the nowcast only slightly overestimates PSD for the rest of the wavelengths from 2 km to 256 km. Even at 3 hours lead time, although the two PSD curves are slightly off at wavelengths greater than 16 km or less than 4 km, it is a near-exact match for wavelengths between 4 km and 16 km, indicating that the forecast contains the same amount of information detail as the MRMS at these spatial scales. On the other hand, the information content in the HRRR does not match the observed QPE at any wavelength.

The performance of NowcastNet was compared against HRRR as well as persistence and advection. Advection is represented by the PySTEPS[25] algorithm (supplementary information section C). For comprehensive evaluation, 30 heavy precipitation events from January 2021 to April 2024 are examined in this study. Among them, the performance of NowcastNet against MRMS, HRRR, persistence, and advection is shown in supplementary figures S1, S2, S3, and S4. In all the events, NowcastNet exhibits a higher degree of similarities to MRMS. In contrast, HRRR encounters challenges in capturing finer details when compared to MRMS. Persistence assumes precipitation intensity and location will remain the same over time, and likewise we see its performance deteriorate over time. On the other hand, the advection model illustrates the movement but fails to capture the intensities of extreme precipitation and produces blurry nowcasts in the wrong locations.

Various metrics were employed to evaluate NowcastNet's predictions against HRRR in these events. These metrics help determine the model's ability to classify and predict the occurrence and intensity of precipitation events, as well as how well the models predict continuous variables related to hydrological processes, such as rainfall amounts and their spatial distribution. The spatial resolution of MRMS and NowcastNet is finer (1 km) than HRRR (3 km). Here, for fair comparison, the MRMS QPEs and NowcastNet forecasts were upscaled to 3 km. Persistence and advection forecasts were also analyzed at 3 km spatial resolution.

The skill score-based metrics included here are probability of detection (POD), false alarm ratio (FAR), critical success index (CSI), F1 Score, Equitable Threat Score (ETS), and Heidke Skill Score (HSS) (figure 4). In this study, we have used 3 thresholds (t) for extreme precipitation events, $t > 0.1$ mm/h, $t > 16$ mm/h, and $t > 32$ mm/h for all categorical skill scores. We have chosen 16 mm/h and 32 mm/h because they are standard benchmarks used to define extreme events in the literature. Figure 4 shows boxplots of all these metrics for different lead times ($T + 1h$, $T + 2h$, $T + 3h$) and for different thresholds (t > 0.1 mm/hr, t > 16 mm/h and t > 32 mm/h) for pixel-wise evaluation. The box plots show interquartile range, with median values and outliers.

For all lead times, all thresholds, and all metrics in Figure 4, NowcastNet outperforms HRRR, with better median scores across the 30 events. NowcastNet's performance deteriorates for longer lead times for the more extreme thresholds, although still technically outperforming HRRR and advection - in these cases HRRR and advection have the worst possible scores and NowcastNet has only slightly better than the worst possible scores. However, for all thresholds at the $T+1h$ lead time and for the t > 0.1 threshold at longer lead times, NowcastNet's superiority over HRRR is clear. For nearly all metrics and thresholds, the quartiles, minimum, and maximum of the NowcastNet scores are also better than the corresponding quartiles, minimum, and maximum of the HRRR scores, indicating that NowcastNet has an advantage for not just the median of the 30 storms but for most of these storms. Aside from superiority over HRRR, we see that NowcastNet performs better than the baseline methods. Overall, this thorough quantitative evaluation emphasizes NowcastNet's effectiveness relative to HRRR and other baseline methods in predicting extreme precipitation events' intensity and location across the forecast intervals under consideration.

Apart from skill scores, we also employed error and correlation based metrics for comprehensive assessment. Figure 5 shows comparison of precipitation forecasts among NowcastNet, HRRR, Persistence, and Advection for 30 heavy precipitation events. Metrics included here are RMSE, Inverse NMSE, Numerical Bias, Normalized Error, and Pearson's Correlation. Metrics are calculated at pixel level, and they provide insights into how well each forecasting method predicts precipitation amounts compared to observed values.

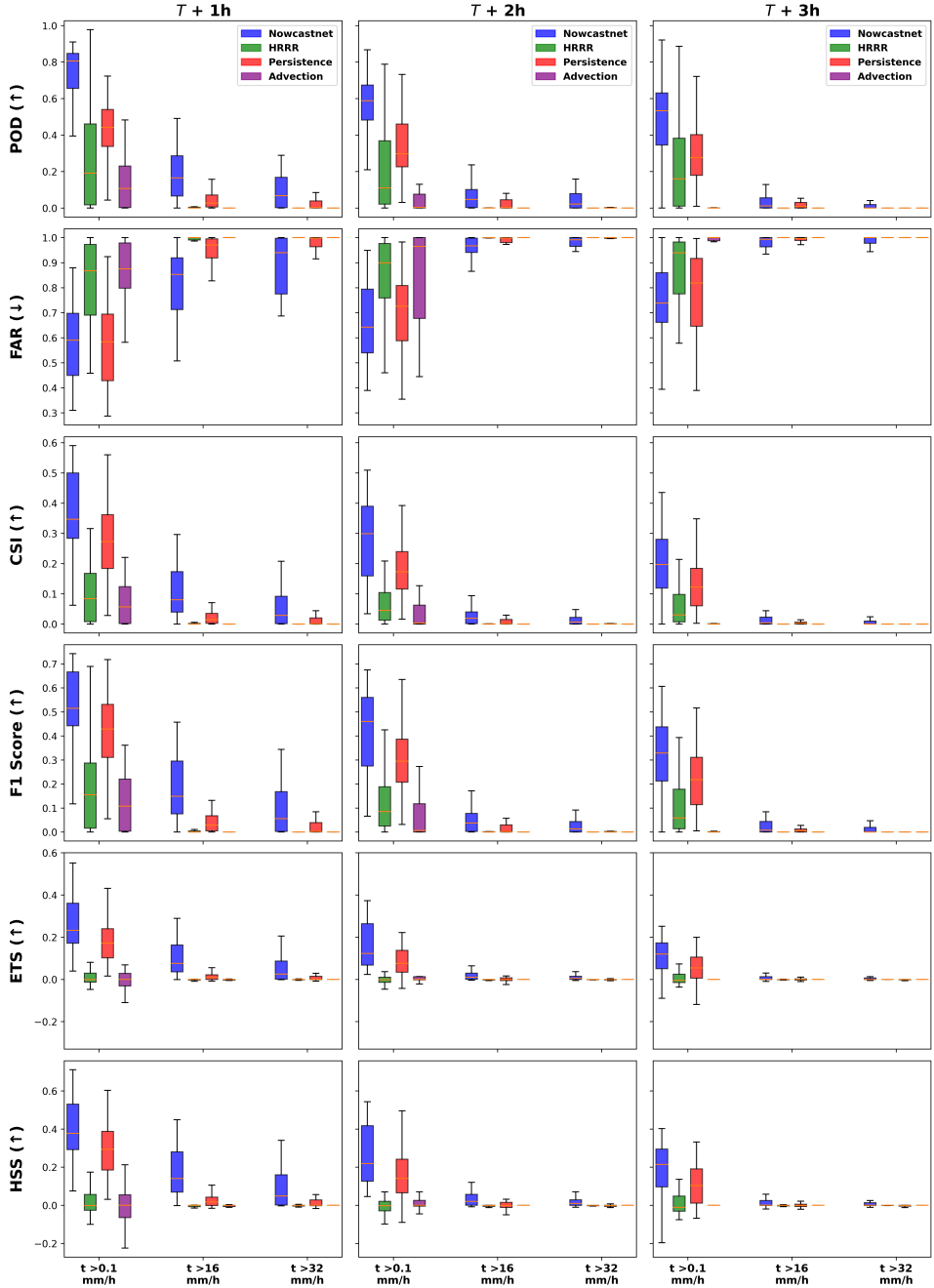

FIG. 4. Comparison of Precipitation Forecast Accuracy between NowcastNet, HRRR, Persistence, and Advection against MRMS QPE (at 3 km spatial resolution) for 30 heavy precipitation events across the geography of interest. Metrics include Probability of Detection (POD), False Alarm Ratio (FAR), Critical Success Index (CSI), F1 Score, Equitable Threat Score (ETS), and Heidke Skill Score (HSS); all at thresholds (t) of t > 0.1 mm/h, t > 16 mm/h and t > 32 mm/h for different lead times (*T*+1h, *T*+2h, and *T*+3h). ↑ indicates higher score is better and ↓ indicates lower score is better.

Findings from these metrics show that NowcastNet outperforms HRRR and baseline methods at 1 hour lead time for all the metrics, indicating its strength in short-term precipitation forecasting accuracy. Also, NowcastNet shows better correlation to MRMS at all lead times than other methods, whereas the median HRRR and advection forecasts show zero correlation at all lead times. However, HRRR and Advection perform better than NowcastNet at longer lead times of 2 hr or 3 hr in terms of RMSE, Inverse NMSE, Numerical Bias and Normalized Error. The reason behind this is NowcastNet's tendency to overestimate more often than other methods, which is shown in its higher positive median Bias across the 30 storms. These results highlight the trade-offs between short-term accuracy and longer-term reliability in precipitation forecasting models, providing insights for improving hydrological predictions during extreme weather events. While NowcastNet is highly effective for immediate predictions, it may sacrifice some accuracy in longer-term forecasts compared to traditional models.

To assess NowcastNet's predictive capabilities at 1 km spatial resolution against MRMS QPE, the set of classification metrics is employed at 10-minute intervals for three thresholds. In figure 6, the metrics are plotted against different lead times. The shaded range in the figure shows the maximum and minimum of 30 heavy precipitation events at any given lead time. The mean POD, indicating the fraction of observed events correctly predicted by the model, ranges from 0.4 to 0.1 for t >16 mm/h and 0.25 to 0.07 for t > 32 mm/h across the evaluated forecasts, and it shows a decreasing trend with increasing lead times. On the other hand, the mean FAR, quantifying the ratio of false alarms to the total number of forecasted events, ranged between 0.62 to 0.97 for both t > 16 mm/h and t > 32 mm/h, with an increasing trend with increasing lead time. The fact that 62%-97% of predicted rain pixels were false alarms is a high FAR, indicating that the predictions can lead to over-warning or unnecessary preparation for precipitation events that do not materialize, potentially causing disruptions or inconvenience to the affected areas. The CSI quantifies the skill of a forecast model in correctly identifying the occurrence and location of specific weather events relative to observed conditions. For an event threshold of 16 mm/h, average CSI values decrease from 0.22 to 0 over time; for an event threshold of 32 mm/h, CSI values decrease from 0.19 to 0.08 over time. The decreasing trend in CSI with increasing lead time suggests that as the forecast lead time extends, the model's ability to accurately predict the location of intense precipitation events diminishes.

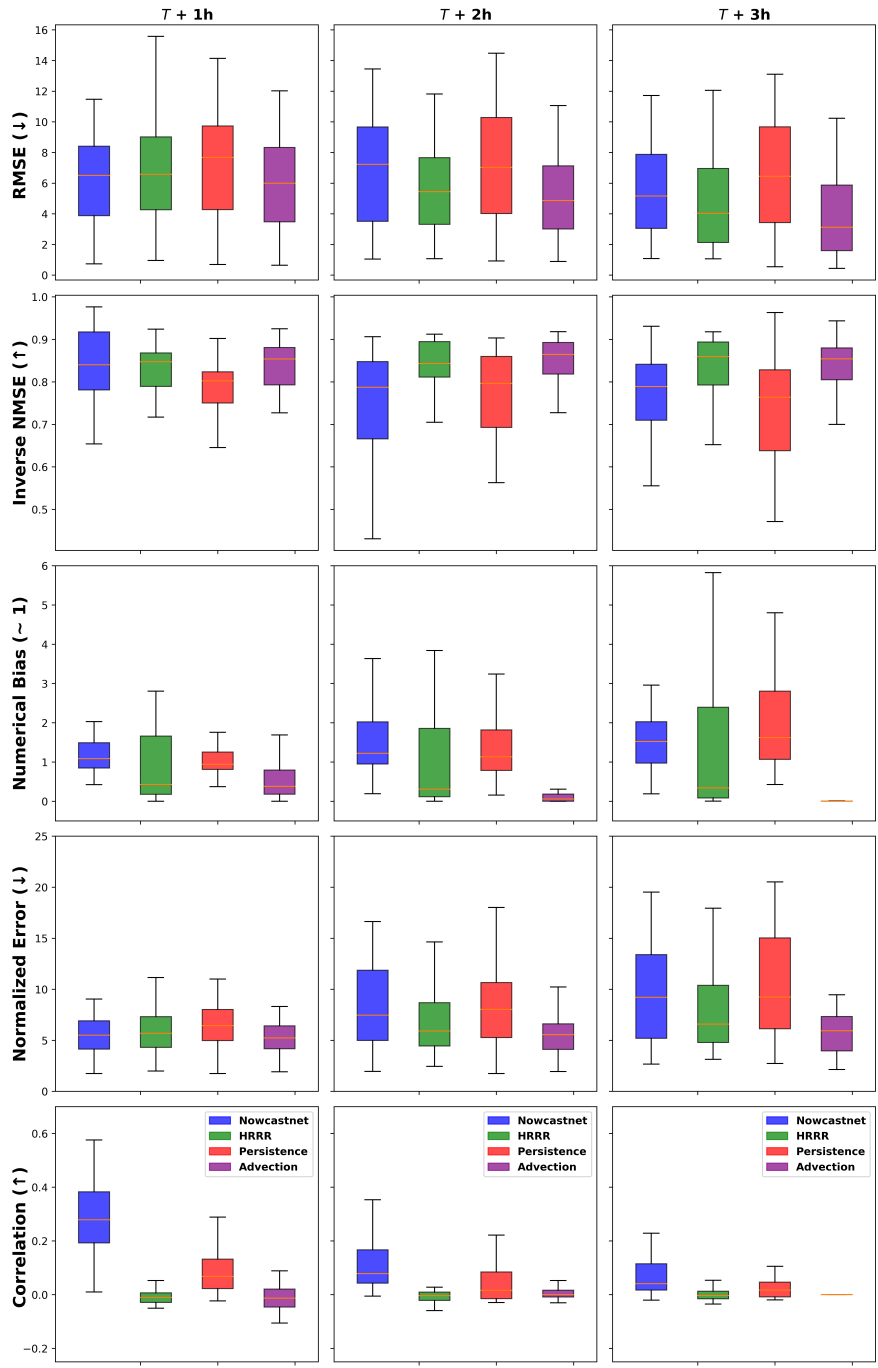

FIG. 5. Comparison of Precipitation Forecast Accuracy between NowcastNet, HRRR, Persistence, and Advection against MRMS QPE (at 3 km spatial resolution) for 30 heavy precipitation events across the geography of interest. Metrics include Root Mean Squared Error (RMSE), Inverse Normalized Mean Squared Error (Inverse NMSE), Numerical Bias, Normalized Error and, Pearson's Correlation for different lead times ($T$+1h, $T$+2h, and $T$+3h). ↑ indicates higher score is better, ↓ indicates lower score is better and ∼ 1 indicates closer to 1 is better.

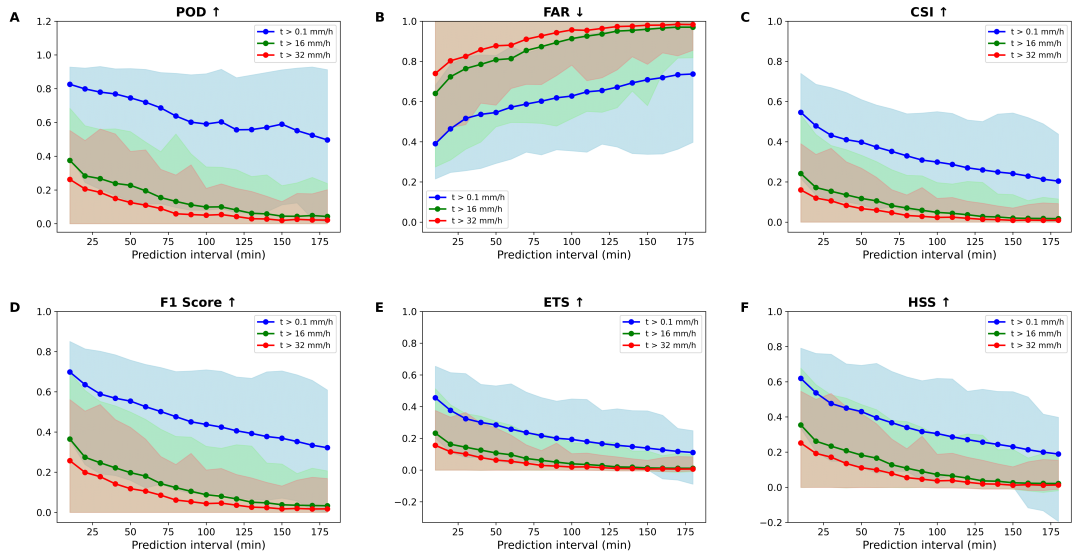

FIG. 6. Performance metrics for evaluating NowcastNet model (at 1 km spatial resolution) with respect to MRMS. Metrics showing (A) Probability of Detection (POD), (B) False Alarm Ratio (FAR), (C) Critical Success Index (CSI), (D) F1 Score (E) Equitable Threat Score (ETS) and (F) Heidke Skill Score; all at thresholds of > 0.1 mm/h, >16 mm/h and >32 mm/h, for MRMS QPE and NowcastNet predictions. ↑ indicates higher score is better and ↓ indicates lower score is better. The shaded region depicts the range between minimum and maximum of 30 heavy precipitation events, and the solid line shows the mean of the 30 events.

The mean F1 score ranges from 0.4 to 0.05 for t > 16 mm/h and 0.25 to 0.02 for t > 32 mm/h over time. The mean value of ETS ranges from 0.2 to 0 and the model's performance is similar at both thresholds; the positive ETS means NowcastNet forecasts can be said to be skilled. The mean HSS value ranges from 0.35 to 0 for a threshold of 16 mm/h and 0.25 to 0 for a threshold of 32 mm/hr. According to the metrics, the model tends to be slightly more successful at delineating events at a 16 mm/h threshold than a 32 mm/h threshold, and for both thresholds this success decreases over time. These trends suggest that as the lead time of the forecast increases, the accuracy and reliability of the model's predictions diminish, leading to larger errors and reduced spatial agreement between forecast and observed precipitation patterns. However, in terms of PSD, the NowcastNet model always performs well when compared to observed QPE, specially at wavelengths of 4 to 16 km (supplementary figures S1, S2, S3 and S4). To make sense of this result in light of the poor CSI, ETS, HSS, and F1 Scores at longer lead times, the explanation is that at a 3-hour lead time, the model is making forecasts with a realistic amount of detail, but not in the correct spatial location.

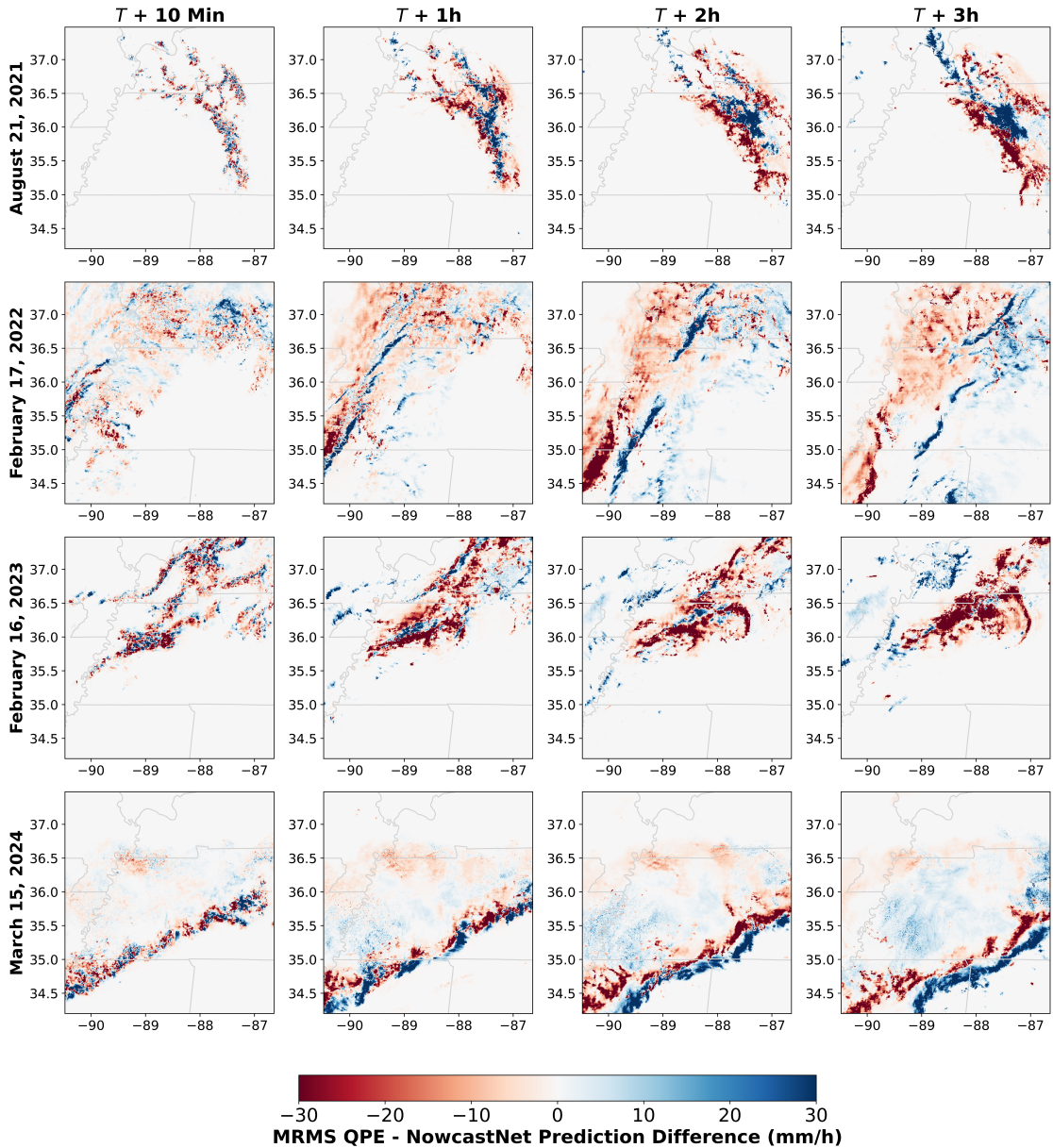

FIG. 7. Comparison of precipitation prediction discrepancies (in mm/h) from NowcastNet model at different lead times ($T$+10 min, $T$ + 1h, $T$ + 2h, and $T$ + 3h) with MRMS for three extreme rainfall events on August 21, 2021 ($T$ = 8:00 UTC), February 17, 2022 ($T$ = 18:00 UTC), February 16, 2023 ($T$ = 12:00 UTC) and March 15, 2024 ($T$ = 3:00 UTC), within the TVA area. The basemap shows US state boundaries. Blue shades represent underestimation, while red shades represent overestimation of precipitation. With increasing lead time, discrepancies between MRMS and NowcastNet predictions become more pronounced.

For further investigation of the NowcastNet model's performance in comparison to MRMS QPE, we assessed four of the heaviest precipitation events from 2021 and 2024. The event forecasts are

done for August 21, 2021 at 8:00 UTC, February 17, 2022 at 18:00 UTC, February 16, 2023 at 12:00 UTC and March 15, 2024 at 3:00 UTC) in the TVA area. We computed the difference between MRMS and NowcastNet predictions for these four events. Figure 7 illustrates a comparison of precipitation prediction discrepancies from the NowcastNet model at different lead times ($T + 10$ min, $T + 1$h, $T + 2$h, and $T + 3$h) with MRMS QPE for the extreme events. The plot shows that, as the lead time increases, discrepancies between MRMS and predictions become more pronounced, indicating the challenges associated with accurately forecasting extreme precipitation events over extended time horizons. We observed areas of underestimation, where NowcastNet either forecast a less-intense event or missed the precipitation event completely, as well as overestimation, where the model predicted high rainfall despite lesser precipitation or lack of it altogether. This suggests that the model's predictive capability diminishes with longer lead times, leading to larger areas of overestimation of precipitation.

## 3. Discussion

Precipitation nowcasting stands as a paramount objective in meteorological science, crucial for informing weather-dependent policymaking. Despite advancements, current numerical weather-prediction systems struggle to provide accurate nowcasts, particularly for extreme precipitation events. In this study, we assessed the efficacy of cutting-edge precipitation nowcasting methodologies, focusing on NowcastNet (a physics conditioned deep generative model) within the TVA service area during extreme precipitation events. NowcastNet's performance was compared against MRMS QPE and HRRR as well as against baseline approaches such as persistence and advection, using various metrics at pixel level with skill score-based metrics such as POD, FAR, CSI, F1 Score, ETS, and HSS, as well as error- and correlation-based metrics such as RMSE, Numerical Bias, Inverse NMSE, Normalized Error, and Pearson's Correlation.

The Waverly event highlighted the challenges of predicting extreme precipitation from mesoscale convective systems. In this event, NowcastNet outperformed HRRR by accurately forecasting hotspots of extreme precipitation over 30 mm/h and predicting the trajectory of convective storms over 3-hour lead times. Also, NowcastNet maintained detailed predictions across most spatial scales, with its spatial power spectral density (PSD) closely matching observed data. In a comprehensive evaluation of 30 heavy precipitation events from 2021 to 2024, NowcastNet consistently showed higher similarity to observed MRMS data compared to HRRR and other benchmarks like persistence and advection, which struggled with fine details and intensity predictions. In terms of skill score and correlation based metrics, NowcastNet's better performance was noticeable against HRRR and other baseline approaches at all lead times and for all thresholds, especially at prediction of extreme precipitation at threshold > 32 mm/h. However, HRRR and Advection surpassed NowcastNet at longer lead times of 2 to 3 hours in terms of RMSE, Inverse NMSE, Numerical Bias, and Normalized Error.

Through the comparison of pixel-based precipitation predictions from the model, areas of both underestimation and overestimation were showcased, the consequences of which are noteworthy. Underestimation can lead to inadequate preparedness and response measures, increasing the risk of property damage, flooding, and even loss of life during extreme weather events. Conversely, overestimation can result in unnecessary disruptions and resource allocation, leading to economic losses and public inconvenience. Therefore, minimizing both underestimation and overestimation is crucial for improving forecast accuracy and enhancing the effectiveness of early warning systems. Notably, NowcastNet's underestimation and overestimation tendencies both intensified with longer lead times. In summary, NowcastNet consistently outperformed HRRR and other models in predicting heavy precipitation events from 2021 to 2024, showing superior similarity to observed MRMS data across various metrics. However, it exhibited shortcomings such as high false alarm rates, inaccuracies in estimating total rainfall, and spatial imprecision at higher resolutions, underscoring the need for continued model refinement.

A salient feature of this study has been the co-evaluation of our nowcasting approach within our team of coauthors consisting of ML developers, hydrologists, water resources engineers and scientists, as well as river managers and hydrometeorologists working at the TVA. The TVA originally discontinued the operational use of HRRR at the request of the river forecast center's (RFC's) lead engineers because it was adding a lot of noise in the early lead times and was inconsistent from run to run. However, they continued examining HRRR predictions as a reference. The inability of this state-of-the-art NWP model, specifically HRRR, to predict extreme rainfall amounts during disastrous flooding events in the TVA region, such as the Waverly event[28], further reinforced their decision to discontinue the operational use of HRRR. A false sense of complacency based on missed predictions of extreme precipitation events, as seemed apparent with HRRR, could lead to inadequate guidance to flooding emergency managers and RFC operators. However, the TVA has remained interested to explore alternatives for improved nowcasting. Based on the results reported here, the physics-embedded ML system, specifically our implementation of NowcastNet, will be evaluated within the operational system of the TVA.

Our research highlights the critical need for further investigations to advance the accuracy of precipitation forecasting. Although deep learning methods have shown promise in nowcasting, compared to baseline methods, a common challenge across these studies is the loss of information content as forecast lead time increases. Also, it has been argued that no method has consistently outperformed Lagrangian persistence (i.e., advection or its variant, optical flow) in improving QPF at scales useful for hydrologic applications, especially for very short lead times (e.g., 1–2 hours)[21;29;30]. On the other hand, while our understanding of the physics behind precipitation, including stratiform and convective rains, continues to advance, translating this knowledge into improved prediction skills, especially at the nowcasting scale, remains challenging. So we hypothesize that by integrating additional physical principles, like momentum conservation, and incorporating diverse ancillary data sources - such as satellite observations, numerical weather predictions, surface observations, land use details, terrain characteristics, and elevation, forecast reliability can be improved. When satellite data are incorporated, the model gains insights into larger-scale weather patterns and atmospheric dynamics, enhancing its ability to capture the spatial and temporal variability of precipitation events. Incorporating information on land use helps the model account for urban effects, vegetation with high transpiration, and bodies of water that increase evaporation, all of which affect precipitation. Terrain properties, including slope, aspect, and roughness, play a noteworthy role in modulating precipitation distribution due to orographic effects and wind patterns. Elevation data further refine precipitation forecasts by capturing changes in atmospheric stability and moisture availability with altitude. Finally, forecasts of state variables from numerical weather predictions (i.e. temperature, air pressure, humidity, wind direction) could be combined with deep generative nowcasts to produce even better forecasts. Figure 8 illustrates the evolution of precipitation forecasting methodologies, showcasing the reduction of information content in forecasts with increasing lead time and highlights the potential of physics-conditioned deep generative models to enhance forecast accuracy through multi source integration and predictive analytics.

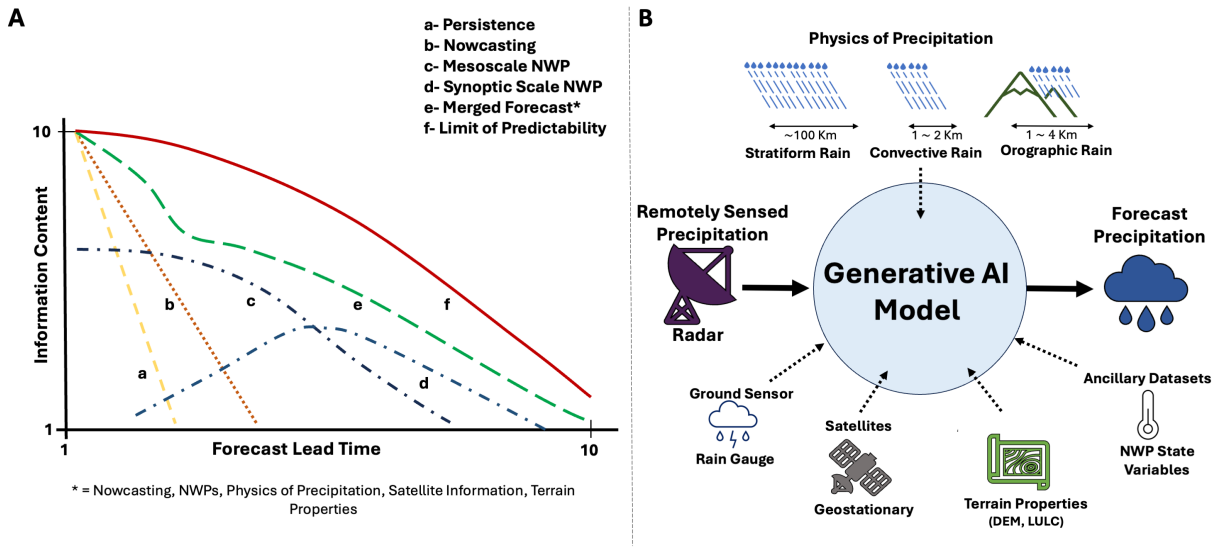

FIG. 8. Multisource Integration and Predictive Analytics in Precipitation Forecasting. Panel A demonstrates the reduction of information content of precipitation forecasts as lead time increases, comparing (a) persistence, (b) nowcasting, (c) mesoscale and (d) synoptic scale numerical weather prediction (NWP), (e) merged approach within the boundary of (f) limit of predictability[31–34]. Merged forecasts can be a combination of nowcasting, NWP models, satellite information etc. Panel B demonstrates generation of precipitation forecasts using a deep generative model (DGM). The proposed DGM combines observed remotely sensed data from radar and geostationary satellites, ground sensors, ancillary information from terrain properties, physics of precipitation[20;35] and NWP state variables to enhance forecast accuracy.

Most of this study's analysis has been done at the grid cell level, but basin-level analysis is important for river management and flood management. The fact that this study reported most metrics at the grid cell level means there has not been quantification of how far away hotspots are when they are wrongly placed - an observed hotspot 2 km away from the forecasted hotspot is much better than 10 km away, so more evaluation is required to understand this. A precipitation hotspot misplaced across basin lines may demand emergency preparations in a completely different river, whereas a precipitation hotspot misplaced within the same basin requires much the same preparations. Geographically incorrect hotspots were a major problem with HRRR, prompting its discontinuation in TVA's decision-making, so basin-wise or dam-wise evaluation of NowcastNet would quantify the confidence that NowcastNet could serve a similar role in hourly-level flood management without the geographic errors.

In conclusion, advancing precipitation nowcasting is crucial for informed decision-making in meteorology, especially for extreme events. While methodologies like NowcastNet show promise in capturing convective events, they exhibit limitations such as false alarms and spatial imprecision. Further model refinement and integration of diverse data sources offer avenues for improvement.

## 4. Methods

*a. Nowcasting Methods*

In this section, we outline the mathematical formulations of various nowcasting techniques, starting with the foundational method of Persistence and progressing through Optical Flow analysis, Numerical Weather Prediction (NWP) models, Machine Learning (ML) techniques, physics-free approaches, and finally physics-conditioned Deep Generative Models.

Persistence-based nowcasting in atmospheric science involves incorporating knowledge of precipitation physics into simple models. Traditional approaches include climatological precipitation history, Eulerian persistence, Lagrangian persistence, and persistence of convective cells[30]. Eulerian persistence (Equation 1) predicts future observations based on the most recent observation, while Lagrangian persistence (Equation 2) accounts for the displacement of air parcels. The Lagrangian persistence assumption is particularly relevant for short-term rainfall prediction and forms the basis of current radar extrapolation models[36].

$$\hat{\psi}(t_0 + \tau, x) = \psi(t_0, x) \tag{1}$$

$$\hat{\psi}(t_0 + \tau, x) = \psi(t_0, x - \lambda) \tag{2}$$

The Eulerian persistence model represents the forecasted precipitation field ($\hat{\psi}$) at a future time ($t_0 + \tau$) as equal to the observed precipitation field ($\psi$) at the initial time ($t_0$), without considering any displacement. In contrast, the Lagrangian persistence model incorporates a displacement vector ($\lambda$) into the equation, representing the movement of air parcels. It forecasts the precipitation field ($\hat{\psi}$) at a future time ($t_0 + \tau$) by shifting the observed precipitation field ($\psi$) at the initial time ($t_0$) by the displacement vector ($\lambda$).

Optical flow techniques, essential in precipitation nowcasting, infer motion patterns from consecutive image frames[19;37]. These methods operate at both local and global scales, utilizing optical flow constraints (OFCs) to delineate motion in specific areas or across entire images[37–39]. Equation (3) describes the Optical Flow Constraint (OFC) equation, which assumes that features within an image sequence maintain their size and intensity while changing shape, serving as the foundation for subsequent models such as STEPS[19].

$$\frac{\delta R}{\delta t} + u\frac{\delta R}{\delta x} + v\frac{\delta R}{\delta y} = 0 \qquad (3)$$

In equation (3), the terms (u,v) represent the velocity field, while R(x,y) denotes the rain rate at the coordinate (x,y). The rain rate R is known at each point, and a sequence of images helps estimate the partial derivatives required in equation (3).

NWP models have improved precipitation forecasting through statistical interpretation, which involves analyzing historical weather data to identify patterns and relationships between various atmospheric variables. However, NWPs only explicitly capture broader weather patterns and so, they are most effective for generating general forecasts 12 hours ahead and beyond[31]. HRRR is an NWP model that played a pivotal role in providing convective storm guidance, especially for aviation meteorologists, over the past decade[23;24;40]. However, with advancements in technology and modeling techniques, the HRRR is transitioning to the Finite Volume Cubed (FV3)-based Rapid Refresh Forecast System (RRFS)[41]. The RRFS represents an evolution from the HRRR, incorporating improvements in resolution, physics parameterizations, and data assimilation techniques[40;42].

In recent years, machine learning has emerged as a promising tool for precipitation nowcasting, offering solutions to limitations in traditional methods like optical flow and numerical weather prediction models (NWPs)[36]. Optical flow methods face challenges due to assumptions of Lagrangian persistence and smooth motion fields, while NWPs struggle to capture fine-scale spatio-temporal patterns associated with convective storms. Machine learning offers potential solutions by capturing complex spatio-temporal patterns, integrating diverse data sources, and introducing approaches like spatiotemporal convolution[16;18;20], adversarial training[21;22;43], and latent random variables[44] to enhance nowcasting capabilities. Among these, the state-of-the-art physics-free deep generative model is DGMR by Google DeepMind[21]. Equation (4) describes the nowcasting methodology of the DGMR model which relies on a conditional generative approach to predict *N* future radar fields based on past *M* observations[21]. This model incorporates latent random vectors *Z* and parameters $\theta$, ensuring spatially dependent predictions by integrating over latent variables[21]. The learning process adopts a conditional generative adversarial network (GAN) framework, tailored specifically for precipitation prediction. Specifically, the model utilizes four consecutive radar observations spanning the previous 20 minutes as contextual input for a generator which enables the generation of multiple future precipitation scenarios over the next 90 minutes[21].

$$P(X_{M+1:M+N}|X_{1:M}) = P(X_{M+1:M+N}|Z, X_{1:M}, \theta)P(Z|X_{1:M})dZ \qquad (4)$$

Although DGMR generates predictions which are spatio-temporally consistent with ground truth for light to medium precipitation events, it produces nowcasts with unnatural motion and intensity, high location error, and large cloud dissipation at increasing lead times[22]. So, in this study, we focus on the state-of-the-art physics conditioned deep generative model NowcastNet[22]. This model employs a physics-conditional deep generative architecture to forecast future radar fields based on past observations, as described in equation (5)[22]. It consists of a stochastic generative network parameterized by $\theta$ and a deterministic evolution network parameterized by $\phi$, allowing for physics-driven generation from latent vectors $z$[22].

$$P(\hat{X}_{1:T}|X_{-T_0:O}, \Phi; \Theta) = \int P(\hat{X}_{1:T}|X_{-T_0:O}, \Phi(X_{-T_0:O}), Z; \theta)P(Z)dZ \qquad (5)$$

This integration enables ensemble forecasting, capturing chaotic dynamics effectively and ensuring physically plausible predictions at both mesoscale and convective scales. The modified 2D continuity equation for precipitation evolution[22], can be represented as:

$$\frac{\delta x}{\delta t} + (\vartheta.\nabla) = s \qquad (6)$$

In this equation, $x$, $\vartheta$, and $s$ represent radar data pertaining to composite reflectivity, motion fields, and intensity residual fields, respectively. The symbol $\nabla$ denotes the gradient operator. This equation represents the conservation of mass for precipitation fields over time and space. In simpler terms, it describes how precipitation changes and moves within a given area, considering factors like radar reflectivity, motion fields (velocity of precipitation movement), and intensity residual fields (changes in precipitation intensity). NowcastNet adaptively combines mesoscale patterns governed by physical laws with convective-scale details from radar observations, resulting in skillful multiscale predictions with up to a 3-hour lead time[22].

*b. Evaluation Metrics*

Evaluation metrics serve as crucial tools for assessing NowcastNet's performance in generating precipitation nowcasts. Murphy described three pillars of forecast evaluation[45]. Firstly, consistency, which refers to the harmony between forecasters' judgments and the forecasts they generate. Secondly, quality, which assesses the concordance between the forecasts and the corresponding observations. Lastly, goodness, which can be thought of as value, evaluates the incremental economic or other benefits realized by decision-makers through the application of the forecasts[45]. We employed a set of metrics to evaluate the performance of NowcastNet and HRRR with respect to MRMS. These metrics include six classification metrics or categorical skill scores: Probability of Detection (POD), False Alarm Ratio (FAR), Critical Success Index (CSI), F1 Score, Equitable Threat Score (ETS) and Heidke Skill Score (HSS). We estimated Power Spectral Density (PSD) for frequency analysis. All of the above scores provide understanding of performance at the fine spatial scales relevant to extreme convective storms. We also employed error and correlation based metrics which include Root Mean Squared Error (RMSE), Numerical Bias, Inverse NMSE, Normalized Spatially Averaged Error, and Pearson Correlation. Together, these metrics provide a comprehensive evaluation of the model's predictive capabilities and reliability against MRMS QPE.

The categorical scores are derived from the 2x2 contingency table, also known as a confusion matrix, clarifying which pixels were observed as events in MRMS and which pixels were forecast as events by the model. Common nomenclature refers to *a* as Hits, *b* as False Alarms, *c* as Misses,

TABLE 1. Contingency Metrics

|  | Forecast Positive | Forecast Negative |
| --- | --- | --- |
| Observation Positive | **Hits** ($a$) | **Misses** ($c$) |
| Observation Negative | **False Alarms** ($b$) | **Correct Negatives** ($d$) |

and *d* as Correct Negatives, Correct Nonevents, or Correct Rejections.

The probability of detection (POD) measures the fraction of observed events correctly predicted by the model (Equation 7) and the false alarm ratio (FAR) quantifies the ratio of false alarms to the total number of forecasted events (Equation 8).

$$POD = \frac{a}{a+c} \quad (7)$$

$$FAR = \frac{b}{a+b} \quad (8)$$

The Critical Success Index (CSI)[46] assesses binary forecasts, determining whether rainfall surpasses a specified threshold t. It provides a comprehensive evaluation of binary classification performance, accounting for both false alarms and misses, and is widely used in the forecasting domain. The CSI measures the ratio of correctly predicted events to the total number of observed and forecasted events (Equation 9).

$$CSI = \frac{a}{a+b+c} \quad (9)$$

CSI is prone to bias because it tends to yield lower scores for rare events[47;48]. To counteract this bias, another scoring method can be utilized to adjust for hits expected by chance. This method is known as the equitable threat score (ETS) or the Gilbert skill score[49]. ETS is calculated using the formula:

$$\text{ETS} = \frac{(a - a_r)}{(a+b+c-a_r)}, \quad \text{where} \quad a_r = \frac{(a+b)(a+c)}{a+b+c+d} \quad (10)$$

The Equitable Threat Score (ETS) spans from -1/3 to 1. When the score falls below 0, it indicates that the chance forecast is favored over the actual forecast, suggesting the forecast lacks skill.

The Heidke skill score (HSS) was originally introduced by Heidke in 1926[50]. It serves as a skill score for categorical forecasts. It is based on the proportion of correct predictions (both Hits and Correct Negatives) but scales according to correct predictions attributable to chance[51].

$$\text{HSS} = \frac{2(ad - bc)}{(a+b)(a+c)+(c+d)(b+d)} \quad (11)$$

This way, HSS ranges from negative infinity to 1. Negative values indicate that the chance forecast outperforms the actual forecast, while 0 indicates no skill, just as good as chance. A perfect forecast achieves an HSS of 1.

The F1 score combines precision and recall, providing a balance between them. Here, precision measures the fraction of predicted events for which the prediction was correct, indicating how correct the model was when it predicted positive cases. Precision is equivalent to 1 - FAR, so a higher precision means lower false alarm ratio. On the other hand, recall, equivalent to the POD, assesses the fraction of positive cases that were correctly identified by the classifier, indicating how correct the model was when an event (positive) was observed. F1 score (Equation 12) is calculated as the harmonic mean of precision and recall (with a threshold of 0.1 mm/h, 16 mm/h and 32 mm/h, for differentiating precipitation events and non-events), indicating the model's accuracy both relative to its own predictions and relative to observed events.

$$F1\ Score = 2\ \frac{Precision \times Recall}{(Precision + Recall)} \quad (12)$$

Where, $Recall = POD = \frac{a}{a+c}$ and $Precision = \frac{a}{a+b} = 1 - FAR$

The RMSE measures the standard deviation of the residuals or the prediction errors. A low RMSE denotes less difference between the observed and predicted values of the variable of interest. RMSE has the same units as the variable.

$$RMSE = \sqrt{\frac{1}{n}\sum_{i=1}^{n}(x_i - y_i)^2} \quad (13)$$

Here, $n$ = number of lead times, $x_i$ = observed mean precipitation and $y_i$ = predicted mean precipitation.

The numerical bias of the forecasts is defined as the ratio of the mean forecast value to the mean observed value across all pixels. A bias value closer to 1 indicates less bias and more accurate forecasts, while values significantly higher or lower than 1 indicate greater bias and less accurate forecasts. Bias higher than 1 means the forecast overestimates observed MRMS precipitation, and bias less than 1 means the forecast underestimates observed MRMS precipitation.

$$Bias = \frac{Mean\ Forecast\ Value}{Mean\ Observed\ Value} \quad (14)$$

The normalized spatially averaged error, or normalized root mean square error, measures the average prediction error as proportion of the spatial mean precipitation. Higher errors indicate lower forecast skill.

$$Normalized\ Error = \frac{RMSE}{Mean\ Observed\ Value} \quad (15)$$

The inverse of the normalized root mean square errors (Inverse NMSE) measures how well the forecast captures the spread of observed pixel values. It is calculated for pixels with nonzero rainfall. A perfect forecast results in Inverse NMSE approaching infinity. For a stationary process, Inverse NMSE equal to 1 indicates that the RMSE equals the standard deviation of the observed pixels, therefore the forecast is as good as predicting the spatial mean in every location. Inverse NMSE less than 1 means the forecast captures the spread worse than a spatial mean prediction.

$$Inverse\ NMSE = \frac{Standard\ Deviation\ of\ Observed\ Data}{RMSE} \quad (16)$$

Pearson's Correlation assesses the similarity between spatial patterns of observed and forecasted precipitation fields.

$$Correlation = \frac{\sum_{i=1}^{n}(x_i - \bar{x})(y_i - \bar{y})}{\sqrt{\sum_{i=1}^{n}(x_i - \bar{x})^2 \sum_{i=1}^{n}(y_i - \bar{y})^2}} \quad (17)$$

Where, $n$ = number of grid cells, $x_i$ = observed precipitation, $y_i$ = predicted precipitation $\bar{x}$ = mean observed precipitation and $\bar{y}$ = mean predicted precipitation.

The spatial power spectral density (PSD)[52;53] characterizes the distribution of precipitation intensities using Fourier transform techniques (Equation 18). This captures the information content - here variance of rain rate - at different spatial scales. Forecasts whose information content matches observations', at all spatial scales, are more desirable. Power spectral density is a function of wavelength. To compute PSD across the geography of interest, first the Fourier transform is computed in each dimension. The Fourier transform has information about different wavelengths, so bins of wavelengths are created and in each bin, the variance of amplitude of the Fourier signal is taken. Below is the formula[53] for the Fourier transform in one dimension. $F(x_j)$ is the Fourier approximation of the signal $y_j$ at each of the $n$ grid cells $x_j$. $L$ is the length (e.g., in kilometers) of the dataset in this dimension. The values of $k$ from 1 through $m$ are the different wavelengths considered. Then, $a_0, a_k, b_k$ are the Euler-Fourier coefficients that define the signal.

$$F(x_j) = \frac{a_0}{2} + \sum_{k=1}^{m} a_k cos(2\pi k \frac{x_j}{L}) + b_k sin(2\pi k \frac{x_j}{L}) \quad (18)$$

*Acknowledgments.* This work was supported by National Aeronautics and Space Administration (NASA) funded project titled "Remote Sensing Data Driven Artificial Intelligence for Precipitation Nowcasting (RAIN)" under Grant 21-WATER21-2-0052 (Federal Project ID: 80NSSC22K1138) from the NASA Water Resources Program within their Earth Science Applications under their Applied Sciences Program. The authors also acknowledge the support from the Northeastern University (NU) focus area Artificial Intelligence for Climate and Sustainability (AI4CaS), which is a part of The Institute for Experiential AI (EAI) at NU and supported by both the NU Roux Institute and the NU Office of the Provost.

*Data availability statement.* The datasets "MRMS" for this study can be found in the NOAA at https://www.nssl.noaa.gov/projects/mrms or contact the MRMS data teams using mrms@noaa.gov. HRRR operational and experimental output are available on the NOAA High Performance Storage System in their standard folder locations for real-time runs.

*Author Contribution.* PD and ARG conceptualized and formulated the problem. PD, AP, and NB performed the experiments and analyzed the results. NB and MH worked closely as stakeholders to co-develop case studies and insights, besides pointing to relevant data. TJV and KD helped with machine learning model evaluation and interpretation. DS and KvW helped develop hydrologic insights. PD and ARG interpreted the results with help from all authors. PD prepared the manuscript primarily with AP and ARG, while all authors helped in revising and editing.

*Competing interests.* The authors declare no competing interests. [1]

---

## A: TVA Case Study

During the 2020 tropical season, there were numerous instances of overpredictions in QPFs. These overestimations led to operational challenges, as TVA implemented precautionary measures only to later face reductions in forecasted rainfall. The fluctuating forecasts necessitated abrupt adjustments in reservoir operations, resulting in potential lost generation value and public perception issues. In August 2020, the Bear Creek system experienced an unexpected flash flood event. Despite QPF projections indicating minimal rainfall (< 0.1 inch), the region received approximately 8 inches of rainfall within a span of 6 hours. The sudden deluge caught both residents and authorities off guard. Similarly, in 2015, heavy rainfall above Chatuge Dam caught TVA off guard. Initially anticipated to be absorbed by existing storage, the rapid development of inflows required swift action. However, mechanical issues with dam equipment complicated efforts to manage rising inflows. The unforeseen challenges underscored the importance of preparedness and equipment reliability in flood mitigation efforts. Another devastating event occurred in July 2013, where underestimations in rainfall forecasts resulted in widespread flooding across the basin. With tributaries already being at Summer pool levels, heavy rainfall exacerbated inflows into the lower main stem. The resulting floods devastated millions of acres of summer soybean and corn crops, highlighting the far-reaching impacts of inaccurate forecasts.

Although HRRR generates hourly forecasts, there may be a delay of 1 to 2 hours before the forecast is available after it is issued. This delay is presumably attributed to various factors, including the computational processing duration for forecast generation on the Weather and Climate Operational Supercomputing System (WCOSS), data transmission duration, network latency, and subsequent downloading and processing times within the TVA system. Moreover, HRRR forecasts have issues with forecast progression (predicting rainfall after it has already occurred) and spatial aggregation errors (misplacing hotspots in incorrect basins). For such reasons, TVA has stopped using the HRRR model operationally. However, when it was utilized, the initial 12 hours of HRRR data were incorporated into coarser QPF datasets obtained from either the Weather Prediction Center (WPC) or the Lower Mississippi River Forecast Center (LMRFC). This was done in the hope of obtaining a more detailed understanding of weather patterns during the early lead times. Unfortunately, forecasters observed inconsistencies in the magnitude and placement of predicted weather phenomena in the early stages of HRRR forecasts. As a result, the HRRR model was removed from the default input dataset. Presently, TVA relies on forecasts from the LMRFC and occasionally from the WPC, which provides predictions with a frequency of every 6 hours. However, the need for hourly or sub-hourly nowcasting capabilities remains critical for effective flood management.

## B: Applying NowcastNet to TVA Case Studies

In this study we have applied the NowcastNet model to evaluate the performance in extreme precipitation events. The first step required to run the pre-trained NowcastNet model is to process the dataset from Multi-Radar Multi-Sensor (MRMS) data. The radar composites encompass the region spanning from 55°N to 20°N in the north-south direction and from 130°W to 60°W in the east-west direction. These composites are organized on a spatial grid of dimensions 3,500 × 7,000. First, we download MRMS data and clip the region extending 5.12° by 5.12° (512 by 512 grid cells) covering our geography of interest for

29 timestamps at 10-minute intervals over 4 h 40 min. To match how the NowcastNet authors processed MRMS precipitation data, we convert them from .grib2 format into PNG image format, and for all the precipitation rate values we add 3, multiply by 10, and convert into 16-bit integers. The set of input PNGs is renamed to 00000-00.png, 00000-01.png, …, 00000-28.png where 00000-08.png is the image for time $T$, with lower-numbered images occurring earlier in time and higher-numbered images occurring later in time. The NowcastNet model expects input PNGs in this format.

Next, the processed MRMS data are used to obtain a nowcast from the model. The 29 PNGs are placed in a folder that NowcastNet expects, and the model is run. It only uses the first 90 minutes of precipitation data (9 images at 10-minute intervals - time $T$ - 80 min through time $T$) as inputs, and generates precipitation map predictions for the next 3 hours (18 images at 10-minute intervals - time $T$ + 10 min through time $T$ + 180 min) as outputs. In an outputs folder, the NowcastNet model returns both MRMS-based QPE files, labeled with timestamps 1 through 27 (corresponding to PNG indices 0 through 26), and prediction files, labeled with timestamps 10 through 27 in the same formats. We set *npy=True* in NowcastNet's *save_plots()* function, telling it to return both PNG infographics, which show rain rate over the entire 5.12° by 5.12° square, and NumPy arrays of rain rate values over a 3.84° by 3.84° center-crop. Metrics are calculated from the 3.84° by 3.84° center-crop square. All figures in this paper also show the same 3.84° by 3.84° geography where latitude boundary is from 33.64° N to 37.48° N and longitude boundary is from 90.48° W to 86.64°W.

## C: PySTEPS for Advection based Precipitation Forecast

This study utilizes PySTEPS to conduct precipitation forecasts with advection. Initially, the Multi-Radar Multi-Sensor (MRMS) dataset is processed, providing high-resolution precipitation rate estimates. In the precipitation forecasting process, a timestep of 10 minutes is set to update forecasts regularly, while 20 ensemble members are employed to account for prediction uncertainties. The Lucas-Kanade optical flow method is selected for estimating precipitation motion between radar images, facilitating the prediction of precipitation evolution. Utilizing the "steps" nowcasting method enables short-term precipitation forecasts based on available data. A seed value of 42 ensures reproducibility of results in random number generation processes. Additionally, a padding size of 10 is specified to mitigate edge effects and maintain forecast integrity.

Supplementary Table S1: List of Heavy Precipitation Events at TVA region
(Latitude Boundary: 33.64° N to 37.48° N ; Longitude Boundary: 90.48° W, 86.64°W)

| Event Date | Forecast Time (UTC) | Event Date | Forecast Time (UTC) |
| --- | --- | --- | --- |
| March 28, 2021 | 5:00 | April 14, 2022 | 3:00 |
| August 21, 2021 | 8:00 | October 13, 2022 | 3:00 |
| March 18, 2021 | 3:00 | February 2, 2022 | 18:00 |
| June 2, 2021 | 12:00 | February 3, 2022 | 18:00 |
| August 31, 2021 | 3:00 | February 16, 2023 | 12:00 |
| December 11, 2021 | 10:00 | August 4, 2023 | 15:00 |
| March 1, 2021 | 3:00 | March 2, 2023 | 6:00 |
| July 1, 2021 | 20:00 | July 19, 2023 | 12:00 |
| October 6, 2021 | 2:00 | August 9, 2023 | 18:00 |
| February 17, 2022 | 18:00 | August 10, 2023 | 9:00 |
| July 7, 2022 | 20:00 | May 12, 2023 | 3:00 |
| February 24, 2022 | 9:00 | January 3, 2023 | 3:00 |
| December 14, 2022 | 9:00 | March 15, 2024 | 3:00 |
| February 22, 2022 | 15:00 | January 24, 2024 | 20:00 |
| January 16, 2022 | 12:00 | January 9, 2024 | 6:00 |

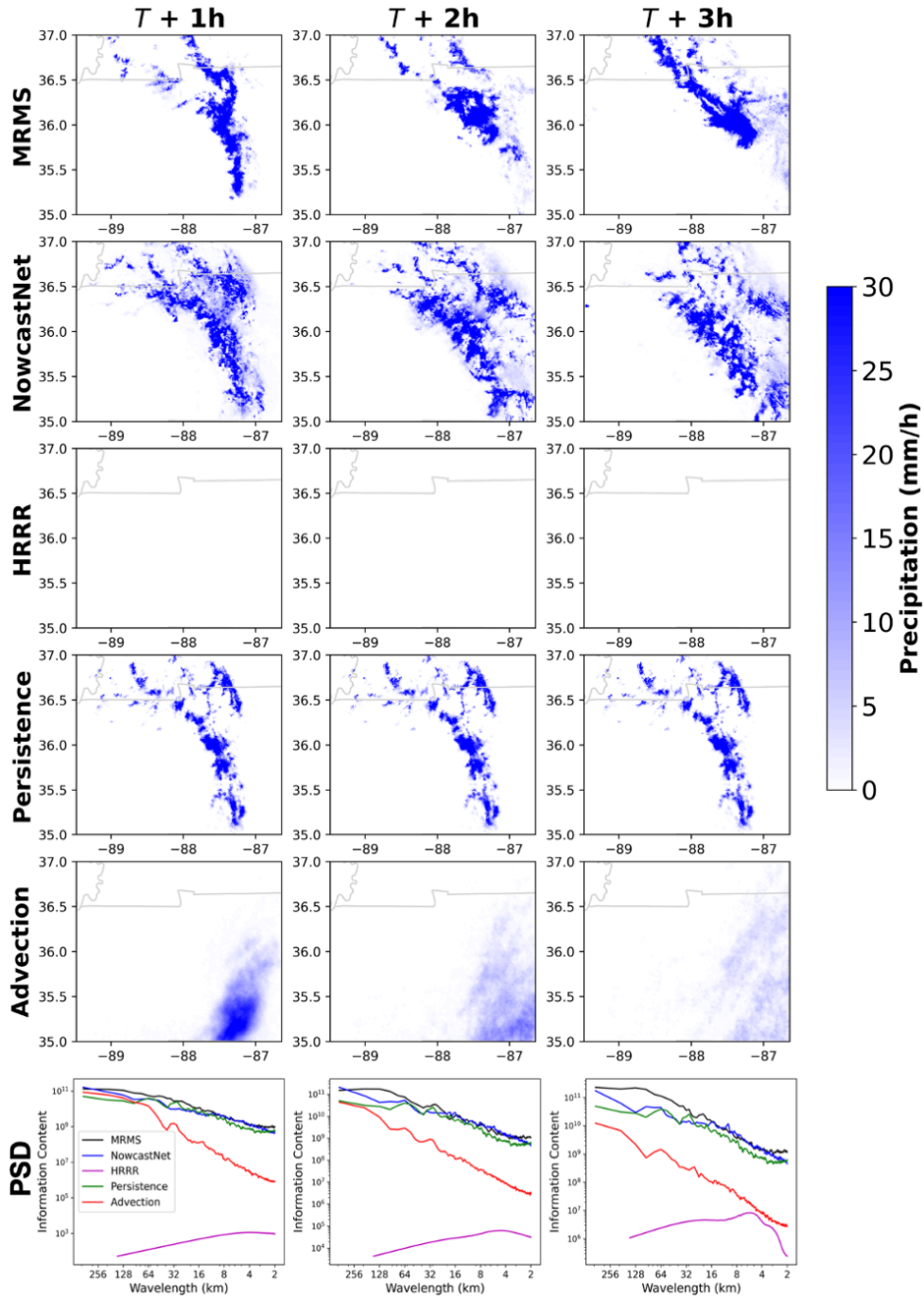

Supplementary Figure S1: Comparison of precipitation forecasts (in mm/hour) from NowcastNet, HRRR, Persistence and Advection at different lead times (*T*+1h, *T*+2h, and *T*+3h) with MRMS QPE for the Waverly flood event on August 21, 2021 (*T* = 08:00 UTC) within the TVA area. The last row depicts the PSD at different wavelengths at different lead times (*T*+1h, *T*+2h, and *T*+3h).

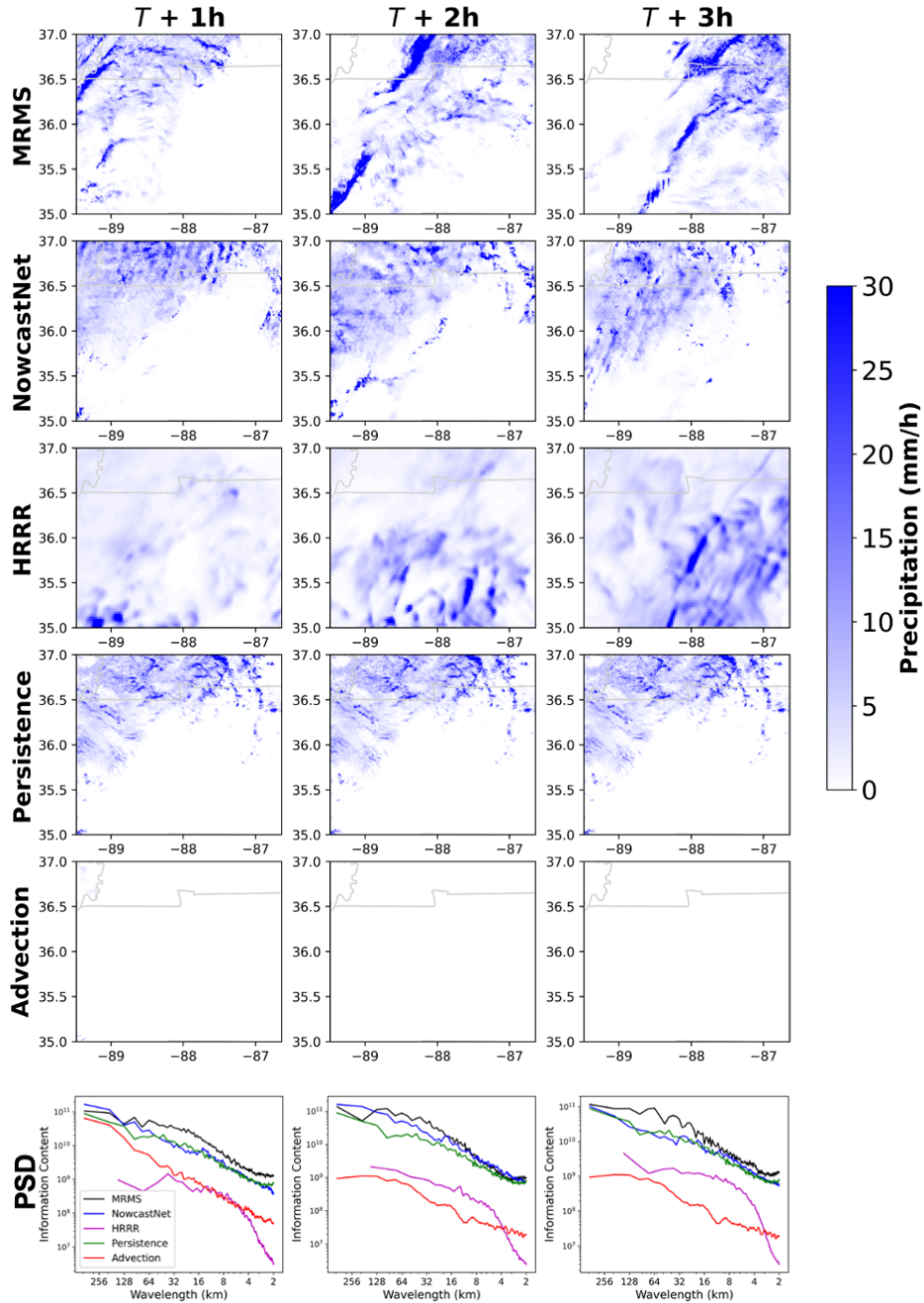

Supplementary Figure S2: Comparison of precipitation forecasts (in mm/hour) from NowcastNet, HRRR, Persistence and Advection at different lead times ($T$+1h, $T$+2h, and $T$+3h) with MRMS QPE on February 17, 2022 ($T$ = 18:00 UTC) within the TVA area. The last row depicts the PSD at different wavelengths at different lead times ($T$+1h, $T$+2h, and $T$+3h).

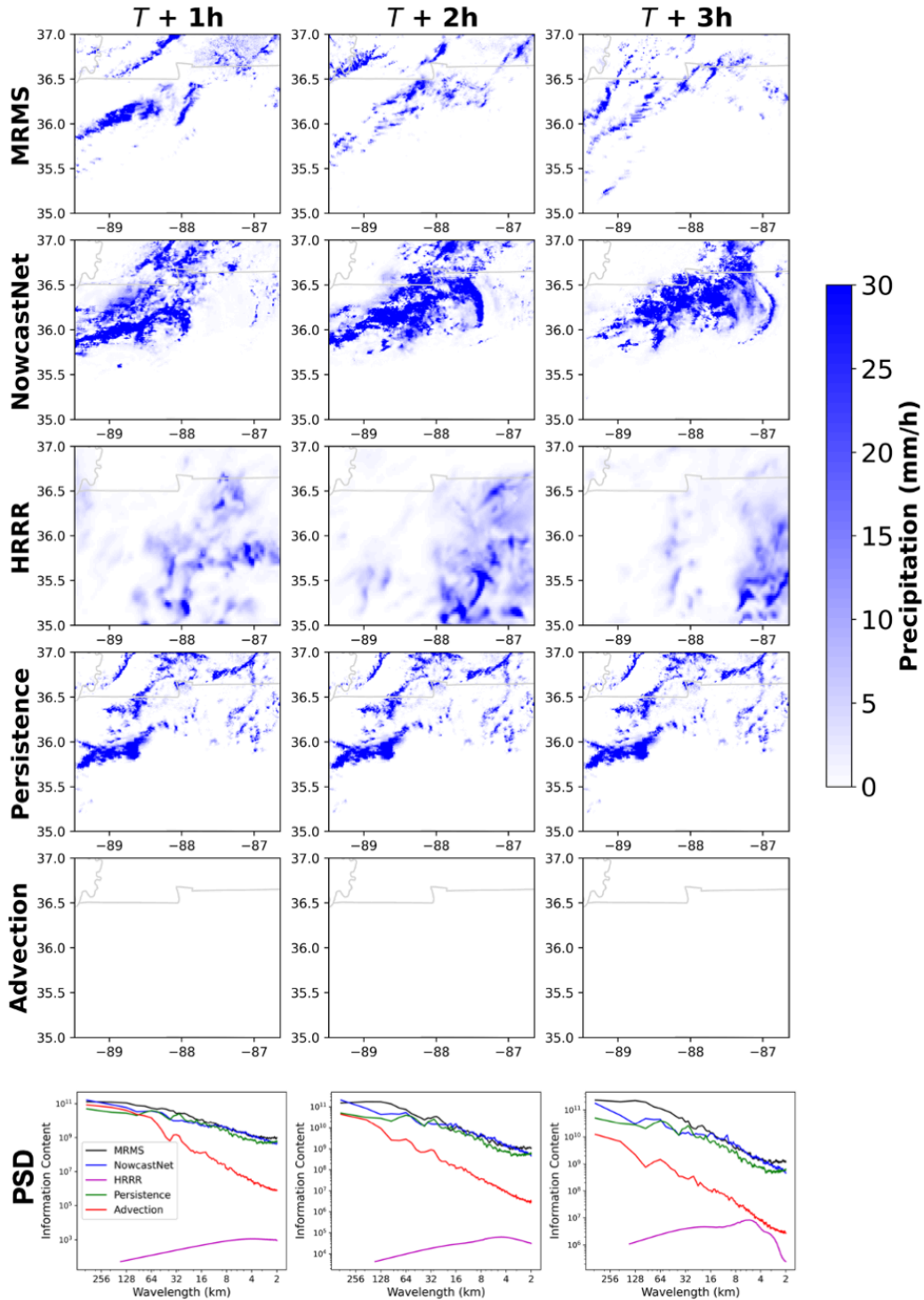

Supplementary Figure S3: Comparison of precipitation forecasts (in mm/hour) from NowcastNet, HRRR, Persistence and Advection at different lead times ($T$+1h, $T$+2h, and $T$+3h) with MRMS QPE on February 16, 2023 ($T$ = 12:00 UTC) within the TVA area. The last row depicts the PSD at different wavelengths at different lead times ($T$+1h, $T$+2h, and $T$+3h).

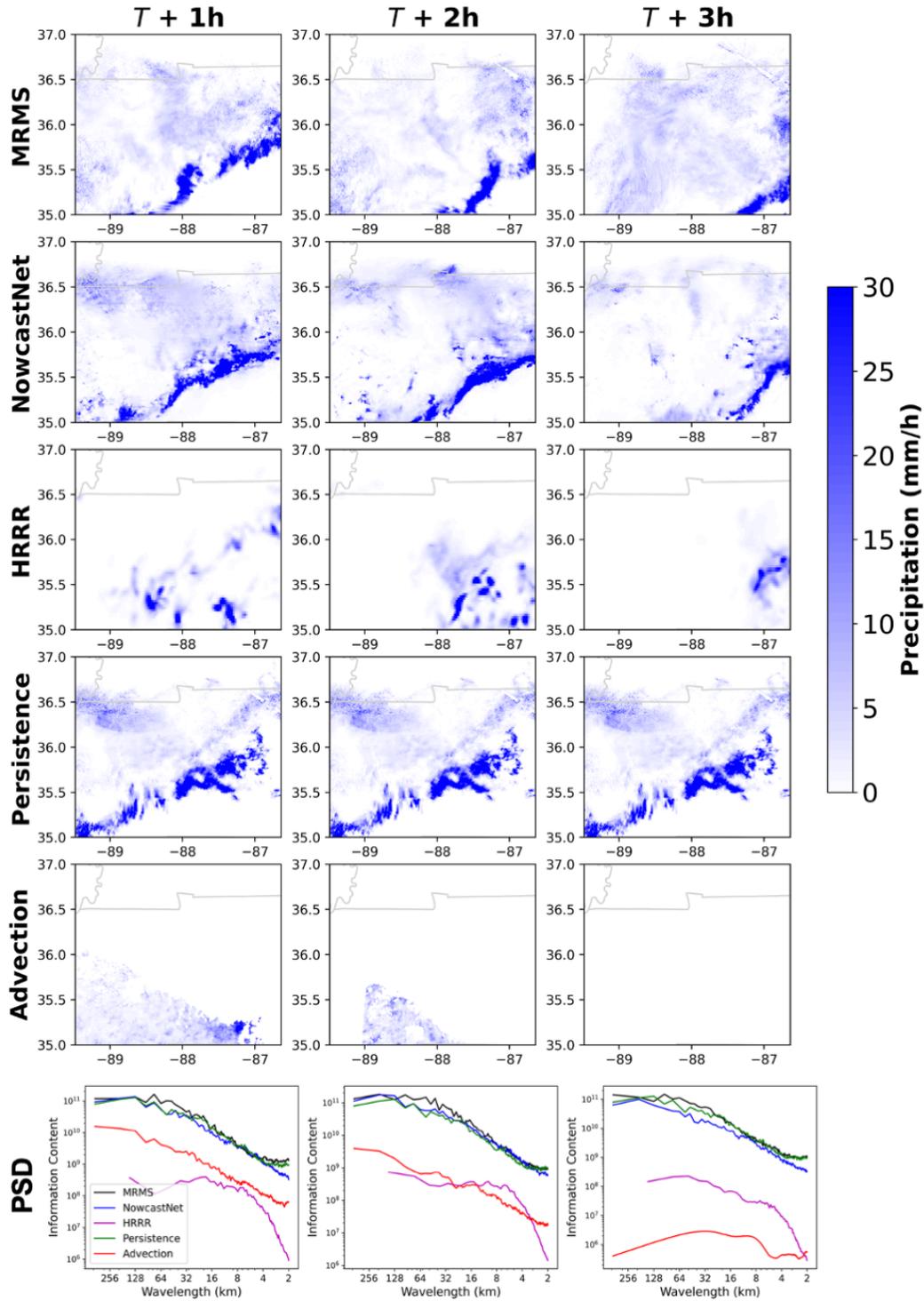

Supplementary Figure S4: Comparison of precipitation forecasts (in mm/hour) from NowcastNet, HRRR, Persistence and Advection at different lead times ($T$+1h, $T$+2h, and $T$+3h) with MRMS QPE on March 15, 2024 ($T$ = 03:00 UTC) within the TVA area. The last row depicts the PSD at different wavelengths at different lead times ($T$+1h, $T$+2h, and $T$+3h).